\documentclass[apj]{emulateapj}
\usepackage{tmsfnts}

\def\astroh     {\emph{Astro-H}}
\def\hitomi     {\emph{Hitomi}}
\def\chandra    {\emph{Chandra}}
\def\xmm        {\emph{XMM}}
\def\xmmnewton  {\emph{XMM-Newton}}
\def\suzaku     {\emph{Suzaku}}
\def\am         {$^\prime$}
\def\kms        {km$\;$s$^{-1}$}
\def\photcms    {phot$\;$s$^{-1}\,$cm$^{-2}$}
\def\cmsq       {cm$^{-2}$}
\def\msun       {$M_{\odot}$}

\submitted{ApJ Letters, submitted 2016 July 24, accepted 2017 February 19}

\begin{document}

\lefthead{{\em HITOMI}\/ CONSTRAINTS ON THE 3.5 KEV LINE IN PERSEUS}
\righthead{{\em HITOMI}\/ COLLABORATION}

\title{{\em HITOMI}\/ CONSTRAINTS ON THE 3.5 KEV LINE IN THE PERSEUS GALAXY
  CLUSTER}

\author{Hitomi Collaboration$^*$:  
F.~A.~Aharonian\altaffilmark{1,2}, H.~Akamatsu\altaffilmark{3},
F.~Akimoto\altaffilmark{4}, S.~W.~Allen\altaffilmark{5,6,7},
L.~Angelini\altaffilmark{8}, K.~A.~Arnaud\altaffilmark{8,9},
M.~Audard\altaffilmark{10}, H.~Awaki\altaffilmark{11},
M.~Axelsson\altaffilmark{12}
% ...
}

\footnotetext[*]{Corresponding authors: M. Markevitch, C. Kilbourne, and
  T.~Tamura (maxim.markevitch@nasa.gov, caroline.a.kilbourne@nasa.gov,
  tamura.takayuki@jaxa.jp). Authors' affiliations are listed at end.}

\begin{abstract}

High-resolution X-ray spectroscopy with \hitomi\ was expected to resolve the
origin of the faint unidentified $E\approx 3.5$ keV emission line reported
in several low-resolution studies of various massive systems, such as
galaxies and clusters, including the Perseus cluster. We have analyzed the
\hitomi\ first-light observation of the Perseus cluster. The emission line
expected for Perseus based on the \xmmnewton\ signal from the large cluster
sample under the dark matter decay scenario is too faint to be detectable in
the \hitomi\ data.  However, the previously reported 3.5 keV flux from
Perseus was anomalously high compared to the sample-based prediction.  We
find no unidentified line at the reported high flux level. Taking into
account the \xmm\ measurement uncertainties for this region, the
inconsistency with \hitomi\ is at a 99\% significance for a broad
dark-matter line and at 99.7\% for a narrow line from the gas. We do not
find anomalously high fluxes of the nearby faint K line or the Ar satellite
line that were proposed as explanations for the earlier 3.5 keV detections.
We do find a hint of a broad excess near the energies of high-$n$
transitions of S {\sc xvi} ($E\simeq 3.44$ keV rest-frame) --- a possible
signature of charge exchange in the molecular nebula and another proposed
explanation for the unidentified line. While its energy is consistent with
\xmm\ pn detections, it is unlikely to explain the MOS signal. A
confirmation of this interesting feature has to wait for a more sensitive
observation with a future calorimeter experiment.

\end{abstract}


\section{INTRODUCTION}
\label{sec:intro}

The nature of dark matter (DM) is one of the fundamental unsolved problems
in physics and astronomy. Direct particle searches in laboratories as well
as searches for electromagnetic signal from celestial objects have been
conducted with no unambiguous detection so far. X-ray observations of DM
concentrations, such as galaxies and clusters, provide a probe for a
particular DM candidate, a sterile neutrino, which is predicted to decay and
emit an X-ray line (Dodelson \& Widrow 1994; Abazajian et al.\ 2001). Early
searches that provided upper limits on line flux (and thus the particle
decay rate) as a function of line energy (which gives the particle mass) are
reviewed, e.g., in Abazajian et al.\ (2012) and Boyarsky et al.\ (2012).

A possible detection was reported by Bulbul et al.\ (2014, hereafter B14),
who found an unidentified line at $E\approx 3.55$~keV in the stacked
spectrum of a large sample of galaxy clusters using \xmmnewton\ EPIC MOS and
pn.  Within their sample was the Perseus cluster (its central region), whose
signal was particularly strong. B14 also reported a detection from Perseus
with \chandra\ at the same energy. Boyarsky et al.\ (2014) reported an \xmm\ 
detection in the outer region of Perseus. Urban et al.\ (2015) and Franse et
al.\ (2016, hereafter F16) detected the line in several regions of Perseus
with \suzaku; however, Tamura et al.\ (2015) did not detect it in the same
\suzaku\ data.  The 3.5 keV line was also reported from other objects, such
as the Galactic Center (Boyarsky et al.\ 2015) and M31 (Boyarsky et al.\ 
2014). Other sensitive searches did not detect a significant line signal
(e.g., from the Milky Way halo, Sekiya et al.\ 2016; Draco dwarf, Ruchayskiy
et al.\ 2016; stacked \suzaku\ clusters, Bulbul et al.\ 2016). Some of the
nondetections were inconsistent with other detections under the decaying DM
hypothesis (in which the line flux must be proportional to the projected DM
mass), most significantly, in a sample of galaxies (Anderson et al.\ 2015).
We also note here that the signal from Perseus reported by \xmm, \chandra\ 
and \suzaku\ was higher than expected given the signal from the rest of the
cluster sample (B14). Astrophysical explanations of the reported line, in
addition to those considered by B14, have also been proposed; a critical
review can be found in F16. An extensive review of the recent observations
is given, e.g., by Iakubovskyi (2015).

As recognized in all previous studies, the above line detections were near
the capability for CCD detectors --- for a $\sim$100 eV resolution, the line
reported from clusters with a $\sim$1 eV equivalent width (EW) is a 1\% bump
above the continuum, easily affected by errors in modeling the nearby atomic
lines and in instrument calibration. A confirmation with a much better
spectral resolution was considered essential. \hitomi, launched in February
2016 and lost in March (Takahashi et al.\ 2014, 2016) after having returned
a groundbreaking spectrum of the Perseus cluster (\hitomi\ Collaboration
2016, hereafter H16), offered us such a possibility.  We present results
from this dataset below.

We use $h=0.7$, $\Omega_\mathrm{m} = 0.3$ and $\Omega_\mathrm{\Lambda} =
0.7$ cosmology. The cluster heliocentric redshift (average for member
galaxies) is 0.0179 (Strubble \& Rood 1999) and the redshift in the CMB
frame is 0.01737, which gives $d_L=75.4$ Mpc and a scale of 21.2 kpc per
1\am. We use the 68\% ($1\sigma$) confidence level for errors unless stated
otherwise.

\section{DATA}
\label{sec:data}

The Perseus cluster was the first-light target for \hitomi, observed early
in the instrument activation phase with the Soft X-ray Spectrometer (SXS;
Kelley et al.\ 2016). SXS is an array of 35 calorimeter pixels with a 4.9
eV FWHM energy resolution (H16), covering a $3'\times 3'$ field of view
(FOV) at the focus of a Soft X-Ray Telescope (SXT; Soong et al.\ 2016). To
maximize statistics, here we coadd the 230 ks Perseus dataset used in H16
and a later 45 ks pointing for a total exposure of 275 ks. The former
dataset is a combination of observations 2 on 2016 February 24-25 and 3 on
March 3-5, both pointed $\sim$1\am\ away from the cluster center, while the
latter (observation 4 on March 6-7) is on-center.  The earliest observation
1 was pointed away from the core and is not included.

For these observations, SXS was still protected from possible contaminants
by the closed gate valve (GV) window. It includes a Be filter that absorbs
soft X-ray photons. At $E=3.5$ keV, the GV window transmission is 1/4 of
that in the normal operation mode, yielding the number of photons equivalent
to about 70 ks of normal observations.

\section{ANALYSIS}
\label{sec:analysis}

To fully utilize the SXS high energy resolution, accurate calibration of
gain (the conversion from the amplitude of the detected signal to photon
energy) for each of its 35 pixels is essential. Unfortunately, the
individual pixel gains were changing during the early part of the mission,
and a contemporaneous gain calibration for the SXS array as planned for
later operations was not available. The procedure that we devised to
calibrate the Perseus data is described in H16. For some of the analysis in
H16, an additional scale factor was applied to force the bright 6.7 keV Fe
He$\alpha$ line from the cluster to appear at the same energy in every
pixel. This additional step removes the true gas velocity gradient across
the cluster along with any residual gain errors. Since DM does not move with
the gas, this would also broaden a DM emission line. However, as reported in
H16, the gas velocity difference across the Perseus core is around 150 \kms,
much less than the expected width of the DM line that we will try to detect.
We use the energy-aligned data in this work, but have confirmed that our
results are essentially the same with or without this final energy-scale
alignment. We do not report the best-fit redshift below because we simply
recover the value used for energy alignment.

We used the Be layer thickness ($270\pm10\;\mu$m) calibrated
using Crab and G21.5--0.9 spectra taken after the Perseus observation.%
\footnote{heasarc.gsfc.nasa.gov/docs/hitomi/calib/hitomi\_caldb\_docs.html}
This differs from the instrument response used in H16 and results in a more
reliable slope of the spectrum in the 3--7 keV band.

The detector energy response (RMF) was generated using the observed energy
resolution of the individual pixels. Its uncertainty is discussed in H16 and
is negligible for this work. We bin the spectrum by 2 eV (which is close to
optimal binning, Kaastra \& Bleeker 2016) and fit using the C statistic
(Cash 1979). The number of counts per 2 eV bin is around 200 in this band,
i.e., the statistics is nearly a Gaussian distribution with
$\sigma=\sqrt{N}$.  The instrumental background is negligible.

\subsection{Systematic uncertainties}
\label{sec:sys}

The SXT has a 1.2\am\ angular resolution (half-power diameter). For our
analysis of the spectrum from the whole $3'\times 3'$ FOV, we do not attempt
to account for PSF scattering in and out of the FOV, and use the instrument
response for an on-axis point source.  We estimate the
effect of this simplification on the model normalization to be $\sim$10\%.

The uncertainty of the Be layer thickness in the GV window, $\pm10\,\mu$m,
corresponds to a $\pm2.5$\% uncertainty for the flux at $E=3.5$ keV.

A more insidious effect may be caused by uncertainty in modeling the SXT
effective area (Kurashima et al.\ 2016). The SXT reflectivity around the Au
M edges was measured on the ground and combined with values from Henke
(1993) for other energies. The ground measurements show $\sim$1\% systematic
deviations from Henke, one of which is in the 3.43--3.68 keV interval above
the Au M1 edge --- at our energies of interest. Given the finite accuracy of
the ground measurements, we consider the possibility that the Henke values
are more accurate. To quantify the effect of this uncertainty, below we will
derive some of the results using both the default area curve (which uses the
\hitomi\ mirror measured reflectivities) and one in which the Henke
values were used above the Au M1 edge. Similar deviations may be seen
  at other Au M edges, but the next one (M2 at 3.15 keV) is well outside our
  interval of interest and we will not consider it.

%%%%%%%%%%%%%%%%%%%%%%%%%%%%%%%%%%%%%%%%%%%%%%
\begin{figure*}
\center
\includegraphics[width=0.9\textwidth, bb=17 195 610 624,clip]%
{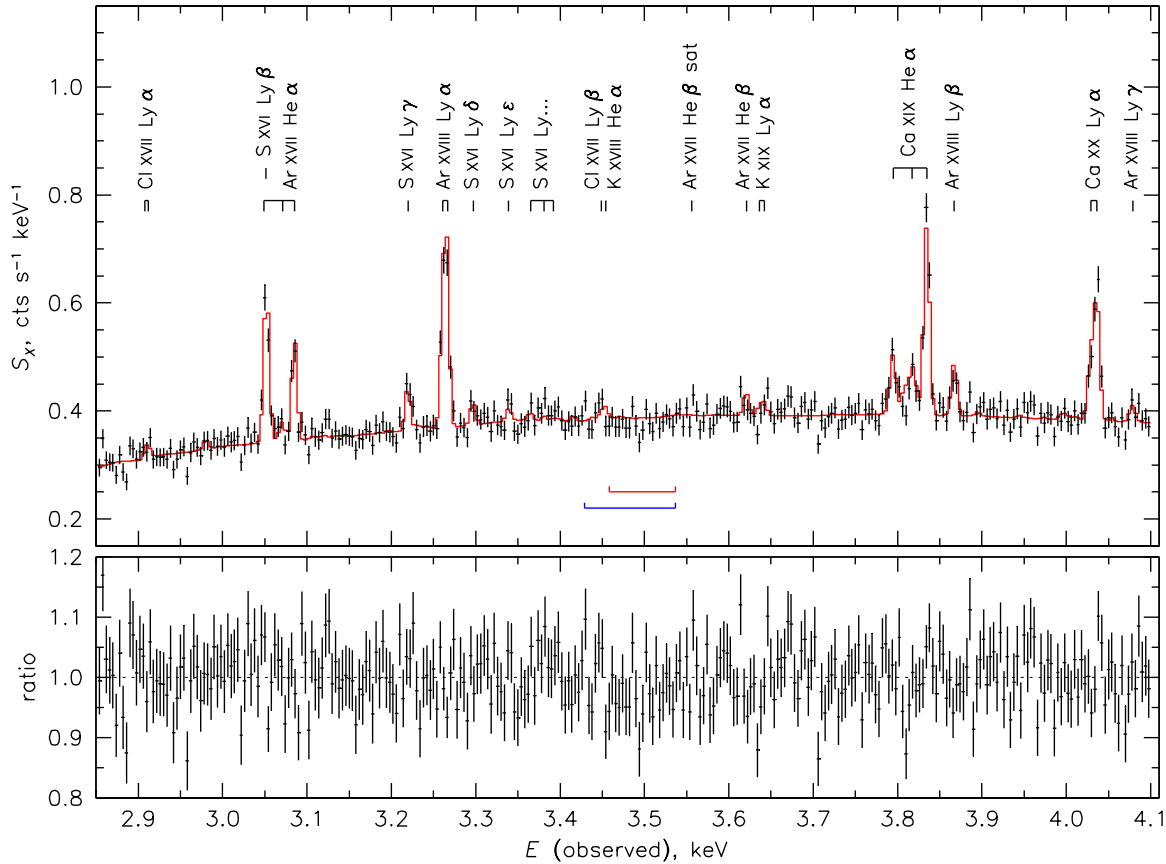}

\caption{SXS spectrum from the whole field of view, combining 3
  pointings. Energy is in the observer frame; bins are 4 eV for clarity (2
  eV bins were used for fitting). Vertical error bars are $1\sigma$ Poisson
  uncertainties in each bin, horizontal error bars denote the bins.  Red
  curve is a best-fit {\sc bapec} model with $kT=3.5$ keV, abundances of 0.54
  solar (same for all elements), l.o.s.\ velocity dispersion of 180 km/s,
  and a power-law component as required by a fit in a broader band (see
  text). Prominent atomic lines seen in the model (identified using AtomDB)
  are marked, along with the interesting Ar {\sc xvii} satellite line (B14)
  that's too faint to be seen in the model. Brackets show 90\% confidence
  intervals on the unidentified 3.5 keV line energy for the most-restrictive
  \xmm\ MOS stacked-clusters sample in B14 (red) and for the \xmm\ MOS
  Perseus spectrum from the region covered by the SXS (blue).}

\label{fig:spec}
\end{figure*}
%%%%%%%%%%%%%%%%%%%%%%%%%%%%%%%%%%%%%%%%%%%%%%

%%%%%%%%%%%%%%%%%%%%%%%%%%%%%%%%%%%%%%%%%%%%%%
\begin{figure}
\vspace*{5mm}
\center
\includegraphics[width=0.47\textwidth, bb=51 184 572 630,clip]%
{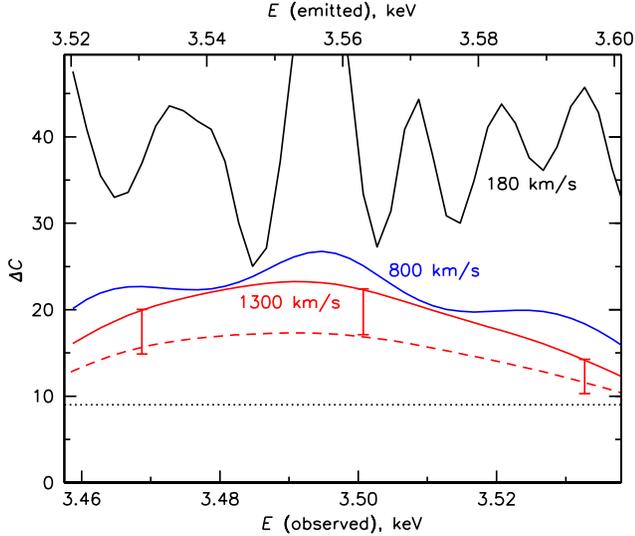}

\caption{Difference of $C$\/ statistic between a model with a line
  with the best-fit \xmm\ MOS flux ($9\times 10^{-6}$ \photcms) and the
  best-fit SXS flux (shown in Fig.\ \ref{fig:limit}; flux is allowed to take
  negative values) as a function of line energy within the B14 most
  restrictive confidence interval for the line. Curves for different line
  widths are shown (black: 180 \kms, blue: 800 \kms, red: 1300 \kms). For
  the 1300 \kms\ case, we also show $\Delta C$\/ between models with the
  \xmm\ MOS line flux and zero line flux (red dashed line). Error bars
  illustrate a systematic uncertainty of the SXT effective area described in
  \S\ref{sec:sys}; its effect is most significant for the broad line and we
  do not show other cases for clarity. Dotted line is at $\Delta C=9$, which
  corresponds to $3\sigma$ exclusion for Gaussian errors.}

\label{fig:excl}
\end{figure}
%%%%%%%%%%%%%%%%%%%%%%%%%%%%%%%%%%%%%%%%%%%%%%

\section{RESULTS}

\subsection{The ICM model}
\label{sec:model}

We fit the full-FOV Perseus spectrum with a {\sc bapec} thermal plasma
model (AtomDB 3.0.3beta2, Foster et al.\ 2012) with elemental
abundances relative to Lodders (2003). We fix the Galactic absorption
at $N_H=1.38\times 10^{21}$ \cmsq\ (Kalberla et al.\ 2005), which
agrees with that derived in the X-ray by \chandra\ (Schmidt et al.\
2002) and \xmm\ (Churazov et al.\ 2003).  A broad-band SXS spectrum
requires a power-law component from the AGN in NGC\,1275 (Fabian et
al.\ 2015). The SXS broad-band effective area calibration is not yet
good enough for fitting multiple continuum components reliably.
Therefore, to derive a spectral shape for the AGN component, we
extracted the AGN spectrum from the off-center \chandra\ Perseus
observations (those where the point-like AGN is not affected by
pileup) and obtained a power-law photon slope $\alpha=-1.8$ (defined
as $S_X\propto E^\alpha$) and an absorption column (Galactic plus
intrinsic) of $3.3\times 10^{21}$ \cmsq. We included a component of
this shape along with the thermal model and fit the SXS spectrum in
the 3--7 keV band, obtaining a normalization for the AGN component of
$9.0\times 10^{-3}$ {phot$\;$cm$^{-2}$\,s$^{-1}$\,keV$^{-1}$} at $E=1$
keV. We fix it in the subsequent fits and leave further discussion of
the AGN spectrum for future work. Its contribution to the 3--4 keV
flux is 15\% and it does not affect our results.

The 2.85--4.1 keV spectrum with the best-fit model is shown in Fig.\ 
\ref{fig:spec}. This energy interval is chosen to include all the
interesting lines but avoid the effective area uncertainty sharply
increasing at lower energies.  The {\sc bapec} model parameters for a fit in
this band are $kT=3.48\pm 0.07$ keV, an abundance of $0.54\pm0.03$ Solar, and
the line-of-sight (l.o.s.)  velocity dispersion of $179\pm16$ km/s (which
becomes $197\pm16$ \kms\ without the pixel energy alignment). The fit is
formally good with C-statistic of 603 ($\chi^2=611$) for 619 d.o.f.  If the
power-law component is omitted, the temperature changes to $3.70\pm0.07$ keV
and abundance to $0.48\pm0.02$.

The parameters obtained from a fit in this narrow interval are qualitatively
similar to those from a broader 3--7 keV band with the power-law slope fixed
at $-1.8$, which gives $kT=3.84\pm0.02$ keV (though with a considerably
higher abundance, $0.68\pm0.01$, now dominated by Fe lines). The closeness
of the best-fit temperatures, even though they are statistically
inconsistent, suggests that the shape of the effective area curve over the
3--7 keV band is reasonably correct. Importantly for this work, the
continuum model at the energy of interest (3.5 keV) differs by only 1\%
between the above fits. We further checked its robustness by fitting a
simple power law in the interval 3.30-3.75 keV between the bright Ar and Ca
lines, excluding intervals with all the weak model lines between, and
obtained a continuum flux only 0.4\% different from our default {\sc bapec}
model. As a further check, we also compared the best-fit normalization of
our {\sc bapec} model to that from \chandra\ for the same region of the
cluster, excluding the AGN. Our normalization is $\sim$10\% below
\chandra's, which is a good agreement, given the preliminary calibration and
the simplified accounting for the PSF.

As seen in Fig.\ \ref{fig:spec}, lines from all elements are fit
surprisingly well with a simple single-temperature, single-abundance
model. Some possible faint lines (K {\sc xviii} He$\alpha$, Ar {\sc
  xvii} He$\beta$, K {\sc xix} Ly$\alpha$) may show problems with line
energies, but none of these lines is a significant detection. Line
identifications and individual abundances will be addressed in a
future work.

\subsection{Constraint on the previously reported 3.5 keV line}
\label{sec:constr}

The red and blue brackets in Fig.\ \ref{fig:spec} show 90\% confidence
intervals for the 3.5 keV line energy for the most sensitive measurement of
B14, that of the \xmm\ MOS stacked-cluster sample, and for the \xmm\ MOS
spectum of the Perseus region covered by \hitomi. For a quantitative
comparison, we extracted a MOS 1+2 spectrum from a circular region
approximating the SXS FOV (both offset and solid angle) in observations 2
and 3 that give most of the exposure, ignoring a small offset for
observation 4.  We then modeled the 3.5 keV line in that spectrum
reproducing the procedure in B14. In particular, we fit the MOS spectrum in
the 2.4--6 keV band using a line-free single-temperature {\sc apec} model
and a set of Gaussian lines at energies of the known atomic lines (with
energies allowed to vary slightly), in order to model the continuum and
lines in as model-independent a way as possible given a CCD detector.  The
faint atomic lines near the energy of interest that could not be directly
detected by the CCD, namely, K {\sc xviii} He$\alpha$ at 3.51 keV
(rest-frame) and Ar {\sc xvii} He$\beta$ satellite at 3.62 keV, were
constrained in the fit using the bright lines of S {\sc xv} He$\alpha$ (2.46
keV rest) and S {\sc xvi} Ly$\alpha$ (2.62 keV), which are good temperature
diagnostics.  The measured S {\sc xv} and S {\sc xvi} fluxes are
$(9.0\pm1.2)\times 10^{-5}$ and $(2.15\pm0.05)\times 10^{-4}$ \photcms,
respectively. A ratio of these lines corresponds to a temperature of 2.9
keV. We predicted the K line flux using this temperature (which is the
relevant one, since K and Ar are likely to come from the same gas phase that
dominates the S lines) and the S line fluxes, assuming the same abundances.
The K {\sc xviii} is a triplet (3.47, 3.49, 3.51 keV) with a known flux
ratio for its components (1:0.5:2.3).  This resulted in an estimate for the
K {\sc xviii} line at 3.51 keV of $2.0\times 10^{-6}$ \photcms. We then
allowed this flux to vary during the fit in the range 0.1--3 times the
estimated flux, capping at $6\times 10^{-6}$ \photcms, to account for
possible temperature and abundance variations. The Ar {\sc xvii} satellite
line is estimated from the measured Ar {\sc xvii} He$\alpha$ line at 3.12
keV, $(6.0\pm 0.3)\times 10^{-5}$ \photcms, and the Ar {\sc xvii}
resonant/satellite line ratio for the above-determined temperature; the
predicted Ar satellite line flux was $2.1\times 10^{-7}$ \photcms, and we
again allowed this flux to vary by factor 0.1--3 in the fit, capping at
$6.3\times 10^{-7}$ \photcms.
  
For the unidentified line, we obtained $f=(9.0\pm 2.9)\times 10^{-6}$
\photcms\ and $E=3.54_{-0.04}^{+0.03}$ keV (similar for the different
assumed line widths from the interesting range). This is very close to the
flux shown in Fig.\ 15 of B14, which gives their \astroh\ prediction, and is
consistent with (but has a much smaller error than) the difference between
the whole-Perseus flux and the one with the central $r=1'$ region excised,
given in their Table 5.
  
We first check how the flux caps for the K {\sc xviii} and Ar {\sc xvii}
satellite lines estimated for the MOS fit compare with the actual fluxes of
those lines in the SXS spectrum. None of the lines is significantly
detected; the flux of the possible blend of K {\sc xviii} He$\alpha$ and
Cl{\sc xvii} Ly$\beta$ (at 3.45 keV observed) is $(4.6\pm2.6)\times 10^{-6}$
\photcms --- under the K {\sc xviii} cap used for the MOS fit.  The Ar {\sc
  xvii} satellite flux (3.556 keV observed) is $(1.5\pm1.4)\times 10^{-6}$
\photcms, consistent with the cap. The above MOS flux of the 3.5 keV feature
is {\em in excess}\/ of these caps, but even if these faint lines were
completely ignored in the MOS fit, neither of them approaches the derived
3.5 keV flux, excluding one of the astrophysical explanations proposed in
B14.
  
The MOS fluxes of the S {\sc xv} He$\alpha$ and S {\sc xvi} Ly$\alpha$
lines, used to derive the K cap, are consistent with the \hitomi\ fluxes,
once the relatively small contribution of Si {\sc xiv} Ly$\gamma$ blending
with S {\sc xv} He$\alpha$ is added. The Ar {\sc xvii} He$\alpha$
MOS-derived flux is consistent with the blend of this line and a $\sim
2\times$ stronger S {\sc xvi} Ly$\beta$ line, resolved in the \hitomi\ 
spectrum (Fig.\ 1); this blending was ingored in the MOS analysis (as in
B14) and resulted in a conservatively high cap on the Ar satellite line. A
more detailed comparison of the \xmm\ and \hitomi\ line fluxes will be given
in a future paper.

We start checking the consistency of the MOS-derived 3.5 keV emission line
with the SXS spectrum by adding a Gaussian line with this flux at a range of
energies to the SXS model. We consider an astrophysical line broadened by
turbulence or a wider line expected from the DM decay. If the astrophysical
line comes from an element whose lines are seen in this range, thermal
broadening would correspond to 100 \kms. Added in quadrature with turbulent
broadening of 180 \kms, this results in an intrinsic Gaussian $\sigma=2.4$
eV at these energies (in addition to the instrumental $\sigma=2.1$ eV, or
4.9 eV FWHM, modeled by the RMF). For a DM line, we try 1300 \kms\ 
($\sigma=15$ eV), which is the l.o.s.\ velocity dispersion of the cluster
galaxies (Kent \& Sargent 1983).  An arbitrary intermediate case of 800
\kms\ corresponds to a lower dispersion in the region of the cD galaxy
projected onto the cluster dispersion. The additional broadening for a
putative DM line caused by our energy alignment (\S\ref{sec:analysis}) is
negligible for such widths, and it would not apply to the narrow line
originating in the gas.

Figure \ref{fig:excl} shows the value of $\Delta C$ (which has the same
interpretation and normalization as $\Delta \chi^2$) for the addition of a
line at the best-fit MOS flux, compared to the best-fit SXS line flux at
that energy (allowing for negative line flux to avoid distorting the
probability distribution, as advised by Protassov et al.\ 2002). For the
broad line, we also show $\Delta C$\/ for a reference model with zero line
flux rather than the best-fit SXS flux.  B14's most restrictive 90\% MOS
energy interval for the stacked sample is shown, since we are assuming that
this is a DM line and it has the same energy in all objects. For narrow and
broad lines, the best-fit \xmm\ MOS flux value is inconsistent with the SXS
spectrum; the weakest constraint is for the broad line and the discrepancy
is at least $\Delta C=12$. Using only observations 2+3 (excluding the
better-centered, but short observation 4) reduced $\Delta C$ for the broad
line compared to the zero-flux model by about 4, commensurate with the
reduction in the number of photons. The effective area uncertainty described
in \S\ref{sec:sys} is illustrated by error bars for the broad line; the
alternative area curve reduces the model values at these energies slightly,
thereby reducing the significance of the exclusion of the \xmm\ flux to at
least $\Delta C=9$. Its effect on the narrower lines is weaker.

\subsubsection{The statistical question}
\label{sec:stat}

To interpret the above $\Delta C$ (or $\Delta \chi^2$) in terms of a
confidence level for the line exclusion, we should note that the statistical
question we are asking --- what is the confidence level of excluding the
previously-detected line --- is different from a blind line search employed
for detecting the line. If a spectral line is detected in a blind search and
it corresponds, e.g., to a $3\sigma$ deviation, one has to estimate the
probability of a false detection under the hypothesis of no line, caused by
a positive random fluctuation. Because a $+3\sigma$ deviation appearing at
{\em any}\/ spectral bin would be detected as a line, such a probability is
the probability of a $+3\sigma$ deviation in one bin times the number of
bins where the line could be found within the searched interval (the
``look-elsewhere'' effect, e.g., Gross \& Vitells 2010; this factor was
applied in B14). However, here we must estimate the probability of a null
hypothesis in which the line exists and we falsely reject it because of a
random negative deviation at the position of the line. While $-3\sigma$
deviations can appear at any spectral bin, only {\em one}\/ of them, that
happens in the bin with the line, would result in false rejection, while all
others would be dismissed as mere random deviations. Thus, even though we do
not know where within the \xmm\ interval the line is, the probability of
false rejection is the probability of a $-3\sigma$ deviation in one bin ---
there is no look-elsewhere effect in our statistical problem. A $\Delta C=9$
or $\Delta \chi^2=9$ corresponds to the standard one-parameter
$1-(1-0.997)/2\approx 99.9$\% confidence level. Because $\Delta C$ is not
constant across the interval in Fig.\ \ref{fig:excl}, we can take its
minimum for a conservative limit for rejecting a certain line flux.

The above $\Delta C$ gives only the \hitomi\ statistical constraint
and does not take into account the fact that the \xmm\ SXS-FOV
detection itself is only $3\sigma$ significant (and thus cannot be
ruled out with a $>3\sigma$ significance). To answer a narrower
question of how inconsistent the \hitomi\ and \xmm\ MOS results for
the same region are, we ran a simple Monte-Carlo simulation with the
line energy and flux randomly drawn from the \xmm\ one-parameter
intervals assuming Gaussian distributions, and the \hitomi\ line flux
at that energy randomly drawn using the \hitomi\ statistical
uncertainty.  For a broad (1300 \kms) line, the SXS line flux was
below the \xmm\ flux in 99.2\% of the trials for the default effective
area, in 98.9\% of the trials if we use the alternative area curve, or
in 97.2\% of the trials if we force the SXS line flux to be zero but
use the same statistical errors. For a narrow (180 \kms) line, for
which the \hitomi\ error is smaller, the discrepancy is at 99.7\% for
all three cases.

%%%%%%%%%%%%%%%%%%%%%%%%%%%%%%%%%%%%%%%%%%%%%%
\begin{figure}
\vspace*{5mm}
\center
\includegraphics[width=0.48\textwidth, bb=50 185 550 650,clip]%
{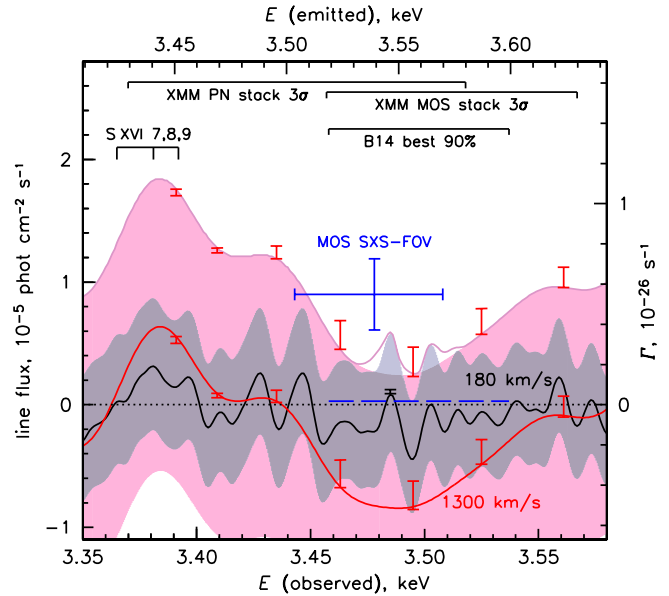}

\caption{The best-fit line flux (solid curves) and the flux limits for
  $C_{\rm min}+9$ ($\pm3\sigma$; shaded bands) for an additional emission
  line as a function of energy. We show an interesting broad band
  encompassing \xmm\ MOS and pn $3\sigma$ intervals for stacked-cluster
  samples from B14 (brackets at top).  Black line with gray band (labeled
  180 \kms) corresponds to a turbulent-broadened line, red line with pink
  band (1300 \kms) corresponds to a DM line. A magenta outline shows
    the highest flux limit from those for different widths in the 180--1300
    \kms\ interval.  Red and black error bars illustrate the systematic
  uncertainty of the effective area (\S\ref{sec:sys}), shown for the
  best-fit curve and the upper limit for the broad line. This effect is
  negligible for the narrow line, so only one location is shown. A line flux
  of $5\times 10^{-6}$ \photcms\ corresponds to EW $\simeq 1$ eV.  Blue
  cross shows the MOS detection for the SXS FOV with $1\sigma$ one-parameter
  uncertainties. Blue dashed line shows the expected flux based on the
    stacked-cluster signal (\S\ref{sec:disc}). Also shown for reference is
  the ``B14 best'' interval covered by Fig.\ \ref{fig:excl}.  The only
  interesting unmodeled positive deviation --- though a low-significance one
  --- is near the energies of the high-$n$\/ transitions of S {\sc xvi},
  marked at top. The right vertical axis shows the approximate corresponding
  sterile neutrino decay rate $\Gamma$.}  \vspace*{3mm}

\label{fig:limit}
\end{figure}
%%%%%%%%%%%%%%%%%%%%%%%%%%%%%%%%%%%%%%%%%%%%%%

%%%%%%%%%%%%%%%%%%%%%%%%%%%%%%%%%%%%%%%%%%%%%%
\begin{figure}
\vspace*{5mm}
\center
\epsscale{1.1}
\includegraphics[width=0.47\textwidth]%
{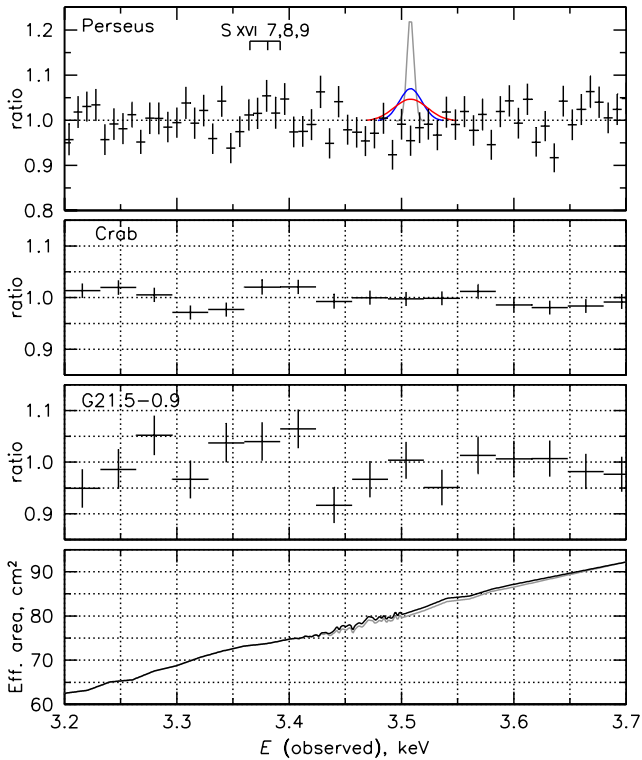}

\caption{Ratios of data to best-fit models in the 
  interesting energy range. {\em Upper panel}\/ shows ratio of the same
  Perseus spectrum and model as in Fig.\ \ref{fig:spec}, but binned by 8 eV.
  A line at 3.57 keV (rest-frame) with a flux derived by \xmm\ in the SXS
  FOV (\S\ref{sec:constr}) is shown with curves of different colors, which
  denote different l.o.s.\ velocity dispersions (gray: 180 \kms, blue: 800
  \kms, red: 1300 \kms, see \S\ref{sec:constr}). Position of the potentially
  interesting S {\sc xvi} feature (\S\ref{sec:344}) is marked. {\em Two
    middle panels}\/ show the residuals for power-law sources Crab and
  G21.5--0.9. The area modification (\S\ref{sec:sys}) is not included. The
  Crab spectrum has sufficient statistics to exclude a significant effective
  area artifact around 3.5 keV. {\em Lower panel}\/ shows the effective area
  curve (gray line shows the modification from \S\ref{sec:sys}), including the
  fine structure above the Au M1 edge measured during ground calibration.}

\label{fig:rat}
\end{figure}
%%%%%%%%%%%%%%%%%%%%%%%%%%%%%%%%%%%%%%%%%%%%%%

%%%%%%%%%%%%%%%%%%%%%%%%%%%%%%%%%%%%%%%%%%%%%%
\begin{figure}
\vspace*{5mm}
\center
\includegraphics[width=0.48\textwidth, bb=50 185 550 650,clip]%
{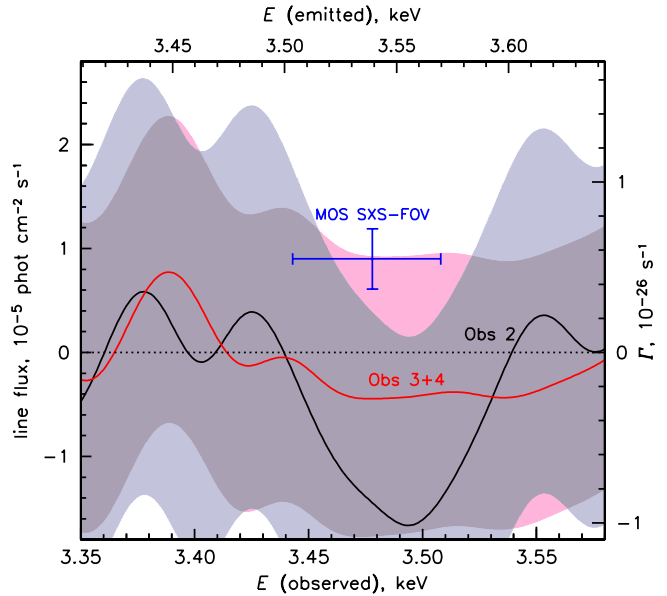}

\caption{The best-fit  flux (curves) and $\pm3\sigma$ limits 
  (shaded bands) for a 1300 \kms\ broadened line (similar to the red line
  and pink band in Fig.\ \ref{fig:limit}), derived separately for the early
  subset (``Obs 2'', black and gray) and later subset (``Obs 3+4'', red and
  pink). The axes and the blue cross are the same as in Fig.\ 
  \ref{fig:limit}. The ``dip'' around 3.5 keV is present in the early subset
  and not in the late one, but the results are statistically consistent.}

\label{fig:limit234}
\end{figure}
%%%%%%%%%%%%%%%%%%%%%%%%%%%%%%%%%%%%%%%%%%%%%%

\subsection{Broader search}
\label{sec:broad}

Figure \ref{fig:limit} shows the best-fit SXS flux for the additional line
as a function of energy, with upper and lower limits at $\Delta C=9$ ($\pm
3\sigma$ for Gaussian distribution), for a narrow and broad line, as well as
the conservative $+3\sigma$ limit selected from among the different line
widths in this range.  The figure shows a wider interval of possible
interest that combines \xmm\ MOS and pn $3\sigma$ line energy intervals. One
notable feature is a broad negative ``dip'' in residuals at $E\approx 3.50$
keV (observed) of about 3--4\%, also noticeable in residuals in Figs.\ 
\ref{fig:spec} and \ref{fig:rat}. The model lines with different widths
overplotted in Fig.\ \ref{fig:rat} show that a broad line may be affected
but not a narrow line, as indeed seen in Fig.\ \ref{fig:limit}. This
deviation has a relatively low statistical significance ($\sim2.3\sigma$).
We have checked the SXS spectra of Crab and G21.5--0.9 (Fig.\ \ref{fig:rat};
details will be given in forthcoming papers), both continuum sources
well-fit with a simple power law model in the energy range of interest.
Neither source shows any comparable deviations at this energy. The Crab
spectrum (shown binned to 32 eV, which roughly corresponds to the expected
DM line width) has 1.5 times more counts at these energies than the Perseus
spectrum, and has sufficient statistics to exclude any effective area
artifact around 3.5 keV of much more than $1\%$ (the size of the errorbars).
The area systematic uncertainty (\S\ref{sec:sys}), shown in the lower panel
of Fig.\ \ref{fig:rat}, is also a smaller (1\%) effect. The fine structure
of the Au M1 edge (same panel), measured with high energy resolution during
ground calibration, occurs on energy scales smaller than the ``dip''. 

We have also checked if this dip may be caused by some time-dependent
instrumental effect. For this, we divided the full Perseus dataset into the
early and late subsets --- observations 2 and 3+4, respectively, separated
by a week (\S\ref{sec:data}). Results from these subsets for the broadened
line, analogous to those shown by red line in Fig.\ \ref{fig:limit} for the
full exposure, are shown in Fig.\ \ref{fig:limit234}. The dip appears in the
early subset but not in the late one. However, the subsets are only
$\sim2\sigma$ apart at 3.5 keV, so the statistics are insufficient to
determine if this is a systematic time-dependent change. The Crab
observation (Fig.\ \ref{fig:rat}) was performed later than our late subset
and thus does not help in ruling out a transient instrumental artifact in
earlier data; however, we can not think of a physical explanation for such
effect. Given the available data, we have to conclude that the dip is most
likely an unfortunate statistical fluctuation, and base our results on the
whole dataset in order to avoid statistical biases.

\subsubsection{A possible excess at 3.44 keV (rest frame)}
\label{sec:344}

The only positive deviation in Fig.\ \ref{fig:limit} is a broad excess above
the best-fit thermal model at $E=3.38-3.39$ keV (observed). The statistical
significance of this feature is only $1.5\sigma$ and it would not be worth
mentioning, if not for the fact that it is located at the energy of the
high-$n$\/ to $n=1$ transitions of S {\sc xvi}. Excess flux in these
transitions can be interpreted as a signature of charge exchange between
heavy nuclei coming in contact with neutral gas --- possibly the molecular
nebula observed in the Perseus core. These particular transitions were
proposed as a possible explanation for the 3.5 keV line in clusters by Gu et
al.\ (2015).  A detection of
charge exchange in the ICM would be of great astrophysical importance, but
it should be confirmed with other elements (to be addressed in future
work) and eventually with a longer exposure.

\section{DISCUSSION}
\label{sec:disc}

Our analysis of the \hitomi\ spectrum of the Perseus cluster core reveals no
unidentified emission line around the energy reported by B14. It is
inconsistent with the presence of a line at the flux reported by B14 using
\xmm\ MOS (as rederived for the approximate SXS FOV). Taking into account
the uncertainties of the \xmm\ MOS measurement in this region, which itself
is only $3\sigma$ significant, the inconsistency with \hitomi\ for a broad
line (that would be emitted by DM) is at the 99\% confidence level, and
99.7\% for a narrow line from the ICM.  The broad line
exclusion level is 97\% if we force the SXS line flux to be zero, assuming
in effect that the mild ``dip'' in the residuals (\S\ref{sec:broad}) is not
statistical as we concluded, but some instrumental artifact present only in
Perseus and not in other SXS data. We note here that F16, using \suzaku\ 
data for a similar Perseus region, reported a line flux and its uncertainty
similar to that from the \xmm\ SXS-FOV measurement, but given the lack of
consensus between different \suzaku\ analyses (cf.\ Tamura et al.), we leave
a comparison with \suzaku\ for a later work.

We can exclude one of the 3.5 keV line astrophysical explanations proposed
by B14 --- namely, anomalously bright K {\sc xviii} He$\alpha$ or Ar {\sc
  xvii} He$\beta$ satellite lines. These lines are not significantly
detected in the SXS spectrum, their fluxes are consistent with expectations
and below the MOS 3.5 keV flux. If we consider a slightly wider energy range
(Fig.\ \ref{fig:limit}), there is a hint of a broad excess emission feature
of the right amplitude (though at very low statistical significance) at
$E\approx 3.44$ keV rest-frame, where charge exchange on S {\sc xvi} has
been predicted (Gu et al.\ 2015).  However, the energy of this feature is
$2.6\sigma$ (100 eV) away from the best-fit energy for the MOS SXS-FOV
detection, and even more inconsistent with the MOS stacked-cluster sample,
though it is consistent with the pn detections (B14). If confirmed with
better statistics, it is an interesting feature in itself.

Given \hitomi's much greater spectral resolution, it is likely that the
inconsistency with \xmm\ that we reported here is attributable to a
systematic error in the \xmm\ result. Possible causes will be examined in a
future work, using the new accurate knowledge of the fluxes of all the
nearby atomic lines from \hitomi, as well as \suzaku\ and \chandra\ spectra
and models. One possible reason, mentioned among the Caveats in B14, is that
with a CCD resolution, a spurious $\sim$1\% dip in the effective area curve
is all that is needed to produce a false line-like residual of the observed
amplitude (see Fig.\ 7 in B14). This is an obvious problem for detections in
a single object or in local objects, even when different instruments with
similar low-resolution detectors are used.  Such systematic effects can be
minimized by stacking objects at different redshifts. In the cluster sample
of B14, the 3.5 keV rest-frame energy spans a 1.2 keV interval of detector
energies, which should smear out any such instrument features. Thus this
systematic error will be much smaller in the stacked-sample signal.

As noted in B14 and subsequent works, the reported line in Perseus, and
especially in its core, was much brighter than expected from the signal in
the larger cluster sample, scaled by mass under the decaying-DM hypothesis.
Assuming that the high Perseus line flux is an artifact but the
stacked-sample signal is real, we can evaluate the corresponding expected
flux from the SXS FOV. To estimate the projected dark matter mass within
this region, we use a total mass profile from Simionescu et al.\ (2011) and
one from the Vikhlinin et al.\ (2006) $M-T$\/ scaling relation (the former
was used in Urban et al.\ and the latter in B14), correcting them for the
14\% baryon fraction. The projected DM mass within the SXS FOV is
$(6-8)\times 10^{12}$\msun. For the sterile neutrino decay rate derived in
B14 for the full cluster sample ($\Gamma\approx 2\times 10^{-28}$ s$^{-1}$),
we expect a 3.5 keV line with $f=(2.4-3.1)\times 10^{-7}$ \photcms\ (see,
e.g., B14 for the equations), 30 times below the flux we ruled out above.
This flux, shown by blue dashed line in Fig.\ \ref{fig:limit}, is below the
statistical noise in the current observation.

The right vertical axis in Fig.\ \ref{fig:limit} shows the sterile neutrino
decay rate that corresponds to the line flux on the left axis, using the
median projected DM mass estimate. The \hitomi\ $3\sigma$ upper limits on
$\Gamma$ are, unfortunately, much higher than many earlier constraints (see,
e.g., B14). This is because of the high X-ray brightness of the ICM in the
Perseus core, the short exposure (combined with the GV attenuation), and the
small SXS FOV.

Our results from this relatively short observation illustrate the dramatic
improvement in sensitivity for narrow features from that of CCD detectors.
However, as expected, the improvement for a putative cluster DM line, which
would have a width of 30--35 eV (FWHM), is less significant. The short
\hitomi\ observation excluded the anomalously bright signal reported from
the Perseus core. However, to test the much weaker stacked-sample detection
(provided it withstands the reevaluation of the systematic uncertainties
after the \hitomi\ result) will require the photon statistics comparable to
that of the CCD stacking studies, or looking at objects where the line is
easier to detect. Among clusters, such objects would be non-cool-core
systems, in which the line EW should be an order of magnitude higher for the
same line flux because of the lower ICM background.  A DM line would be
narrower, giving a calorimeter a greater leverage, in systems with low
velocity dispersion, such as dwarf spheroidals and the Milky Way. Of course,
distinguishing a DM line from an astrophysical one would require resolving
the line, which only a calorimeter can do.

\vspace{1cm}
\section*{Acknowledgments}

We are grateful to the referee for insightful comments and insistence that
improved the paper. We thank the JSPS Core-to-Core Program for support. We
acknowledge all the JAXA members who have contributed to the \astroh\ 
(\hitomi) project.  All U.S.\ members gratefully acknowledge support through
the NASA Science Mission Directorate. Stanford and SLAC members acknowledge
support via DoE contract to SLAC National Accelerator Laboratory
DE-AC3-76SF00515 and NASA grant NNX15AM19G.  Part of this work was performed
under the auspices of the U.S.\ DoE by LLNL under Contract DE-AC52-07NA27344
and also supported by NASA grants to LLNL. Support from the European Space
Agency is gratefully acknowledged. French members acknowledge support from
CNES, the Centre National d'Etudes Spatiales. SRON is supported by NWO, the
Netherlands Organization for Scientific Research. Swiss team acknowledges
support of the Swiss Secretariat for Education, Research and Innovation SERI
and ESA's PRODEX programme.  The Canadian Space Agency is acknowledged for
the support of Canadian members.  We acknowledge support from JSPS/MEXT
KAKENHI grant numbers 15H02070, 15K05107, 23340071, 26109506, 24103002,
25400236, 25800119, 25400237, 25287042, 24540229, 25105516, 23540280,
25400235, 25247028, 26800095, 25400231, 25247028, 26220703, 24105007,
23340055, 15H00773, 23000004 15H02090, 15K17610, 15H05438, 15H00785, and
24540232.  H.  Akamatsu acknowledges support of NWO via Veni grant.  M.
Axelsson acknowledges JSPS International Research Fellowship. C. Done
acknowledges STFC funding under grant ST/L00075X/1.  P. Gandhi acknowledges
JAXA International Top Young Fellowship and UK Science and Technology
Funding Council (STFC) grant ST/J003697/2. A. C. Fabian, C. Pinto and H.
Russell acknowledge support from ERC Advanced Grant Feedback 340442. N.
Werner has been supported by the Lend\"ulet LP2016-11 grant from the
Hungarian Academy of Sciences. We thank contributions by many companies,
including, in particular, NEC, Mitsubishi Heavy Industries, Sumitomo Heavy
Industries, and Japan Aviation Electronics Industry.

Finally, we acknowledge strong support from the following engineers.
JAXA/ISAS: Chris Baluta, Nobutaka Bando, Atsushi Harayama, Kazuyuki Hirose,
Kosei Ishimura, Naoko Iwata, Taro Kawano, Shigeo Kawasaki, Kenji Minesugi,
Chikara Natsukari, Hiroyuki Ogawa, Mina Ogawa, Masayuki Ohta, Tsuyoshi
Okazaki, Shin-ichiro Sakai, Yasuko Shibano, Maki Shida, Takanobu Shimada,
Atsushi Wada, Takahiro Yamada; JAXA/TKSC: Atsushi Okamoto, Yoichi Sato,
Keisuke Shinozaki, Hiroyuki Sugita; Chubu U: Yoshiharu Namba; Ehime U: Keiji
Ogi; Kochi U of Technology: Tatsuro Kosaka; Miyazaki U: Yusuke Nishioka;
Nagoya U: Housei Nagano; NASA/GSFC: Thomas Bialas, Kevin Boyce, Edgar
Canavan, Michael DiPirro, Mark Kimball, Candace Masters, Daniel McGuinness,
Joseph Miko, Theodore Muench, James Pontius, Peter Shirron, Cynthia Simmons,
Gary Sneiderman, Tomomi Watanabe; Noqsi Aerospace Ltd: John Doty; Stanford
U/KIPAC: Makoto Asai, Kirk Gilmore; ESA (the Netherlands): Chris Jewell;
SRON: Daniel Haas, Martin Frericks, Philippe Laubert, Paul Lowes; U.\ of
Geneva: Philipp Azzarello; CSA: Alex Koujelev, Franco Moroso.

\section*{Authors' affiliations}

{\small

$^{1}$ {Astronomy and Astrophysics Section, Dublin Institute
  for Advanced Studies, Dublin 2, Ireland} 

$^{2}$ {National Research Nuclear University (MEPHI), 115409,
  Moscow, Russia} 

$^{3}$ {SRON Netherlands Institute for Space Research, Utrecht, The
  Netherlands} 

$^{4}$ {Department of Physics, Nagoya University, Aichi 464-8602,
  Japan} 

$^{5}$ {Kavli Institute for Particle Astrophysics and 
  Cosmology, Stanford University, CA 94305, USA} 

$^{6}$ {Department of Physics, Stanford University, Stanford, CA
  94305, USA}  

$^{7}$ {SLAC National Accelerator Laboratory, Menlo Park, CA 94025,
  USA}  

$^{8}$ {NASA/Goddard Space Flight Center, Greenbelt, MD 20771, USA}

$^{9}$ {Department of Astronomy, University of Maryland, 
  College Park, MD 20742, USA}

$^{10}$ {Universit\'e de Gen\`eve, 1211 Gen\`eve 4, Switzerland}

$^{11}$ {Department of Physics, Ehime University, Ehime 790-8577,
  Japan} 

$^{12}$ {Department of Physics, Tokyo Metropolitan University,
  Tokyo 192-0397, Japan} 

$^{13}$ {Department of Physics, University of Tokyo, Tokyo
  113-0033, Japan} 

$^{14}$ {Kavli Institute for Astrophysics and Space Research,
  Massachusetts Institute of Technology, Cambridge, MA 02139, USA} 

$^{15}$ {Smithsonian Astrophysical Observatory, Cambridge, MA
  02138, USA}  

$^{16}$ {Lawrence Livermore National Laboratory, Livermore, CA
  94550, USA} 

$^{17}$ {Institute of Astronomy, Cambridge University, 
  Cambridge, CB3 0HA, UK}

$^{18}$ {Yale Center for Astronomy and Astrophysics, Yale
  University, New Haven, CT 06520, USA} 

$^{19}$ {Department of Physics, University of Durham, Durham,
  DH1 3LE, UK }

$^{20}$ {Institute of Space and Astronautical Science,
  Japan Aerospace Exploration Agency (JAXA), Sagamihara, Kanagawa 252-5210,
  Japan}  

$^{21}$ {Department of Astronomy, Kyoto University, Kyoto 606-8502,
  Japan} 

$^{22}$ {The Hakubi Center for Advanced Research, Kyoto
  University, Kyoto 606-8302, Japan} 

$^{23}$ {Faculty of Mathematics and Physics, Kanazawa
  University, Ishikawa 920-1192, Japan} 

$^{24}$ {Department of Physical Science, Hiroshima
  University, Hiroshima 739-8526, Japan } 

$^{25}$ {Fujita Health University, Toyoake, Aichi 470-1192, Japan}

$^{26}$ {Physics Department, University of Miami, Coral
  Gables, FL 33124, USA}

$^{27}$ {Department of Astronomy and Physics, Saint Mary's
  University, Halifax, NS B3H 3C3, Canada } 

$^{28}$ {Department of Physics and Astronomy, University of
  Southampton, Highfield, Southampton, SO17 1BJ, UK} 

$^{29}$ {IRFU/Service d'Astrophysique, CEA Saclay, 91191
  Gif-sur-Yvette Cedex, France} 

$^{30}$ {European Space Agency (ESA), European Space
  Astronomy Centre (ESAC), Madrid, Spain} 

$^{31}$ {Department of Physics and Astronomy, Aichi
  University of Education, Aichi 448-8543, Japan} 

$^{32}$ {Department of Physics, University of Maryland,
  Baltimore County, Baltimore, MD 21250, USA} 

$^{33}$ {Department of Applied Physics and Electronic
  Engineering, University of Miyazaki, Miyazaki 889-2192, Japan} 

$^{34}$ {Department of Earth and Space Science, Osaka
  University, Osaka 560-0043, Japan} 

$^{35}$ {Department of Physics, School of Science and
  Technology, Kwansei Gakuin University, 669-1337, Japan} 

$^{36}$ {Department of Physics, Rikkyo University, Tokyo
  171-8501, Japan} 

$^{37}$ {Department of Physics and Astronomy, Rutgers
  University, Piscataway, NJ 08854, USA} 

$^{38}$ {Faculty of Human Development, Kobe University, Hyogo
  657-8501, Japan} 

$^{39}$ {Kyushu University, Fukuoka 819-0395, Japan}

$^{40}$ {Research Institute for Science and Engineering,
  Waseda University, Tokyo 169-8555, Japan} 

$^{41}$ {Department of Physics, Chuo University, Tokyo 112-8551,
  Japan} 

$^{42}$ {Tsukuba Space Center (TKSC), Japan Aerospace
  Exploration Agency (JAXA), Ibaraki 305-8505, Japan} 

$^{43}$ {Department of Physics, Tokyo Institute of
  Technology, Tokyo 152-8551, Japan} 

$^{44}$ {Department of Physics, Toho University, Chiba 274-8510,
  Japan} 

$^{45}$ {Department of Physics, Tokyo University of Science, Chiba
  278-8510, Japan} 

$^{46}$ {Department of Physics, Kyoto University, Kyoto 606-8502,
  Japan} 

$^{47}$ {Universities Space Research Association, Columbia, MD
  21046, USA}  

$^{48}$ {Department of Electronic Information Systems,
  Shibaura Institute of Technology, Saitama 337-8570, Japan} 

$^{49}$ {Space Telescope Science Institute, Baltimore, MD 21218, USA}

$^{50}$ {European Space Agency (ESA), European Space Research
  and Technology Centre (ESTEC), 2200 AG Noordwijk, The Netherlands} 

$^{51}$ {RIKEN, Saitama 351-0198, Japan}

$^{52}$ {Kobayashi-Maskawa Institute, Nagoya University, Aichi
  464-8602, Japan} 

$^{53}$ {Department of Physics, Tokyo University of Science, Tokyo
  162-8601, Japan} 

$^{54}$ {Department of Physics, University of Wisconsin,
  Madison, WI 53706, USA}

$^{55}$ {University of Waterloo, Ontario N2L 3G1, Canada}

$^{56}$ {Department of Astronomy, University of Michigan, Ann
  Arbor, MI 48109, USA}

$^{57}$ {Okinawa Institute of Science and Technology Graduate
  University, Okinawa 904-0495, Japan } 

$^{58}$ {Department of Information Science, Faculty of
  Liberal Arts, Tohoku Gakuin University, Miyagi 981-3193, Japan} 

$^{59}$ {Department of Physics, Faculty of Science, Yamagata
  University, Yamagata 990-8560, Japan} 

$^{60}$ {RIKEN Nishina Center, Saitama 351-0198, Japan}

$^{61}$ {Department of Physics, Faculty of Science, Nara
  Women's University, Nara 630-8506, Japan} 

$^{62}$ {Department of Teacher Training and School Education,
  Nara University of Education, Takabatake-cho, Nara 630-8528, Japan} 

$^{63}$ {The Frontier Research Institute for
  Interdisciplinary Sciences, Tohoku University, Sendai, Miyagi
  980-8578, Japan} 

$^{64}$ {Research Center for Nuclear Physics (Toyonaka),
  Osaka University, 1-1 Machikaneyama-machi, Toyonaka, Osaka 560-0043,
  Japan} 

$^{65}$ {NASA/Marshall Space Flight Center, Huntsville, AL 35812,
  USA}  

$^{66}$ {Department of Astronomy, Columbia University, New
  York, NY 10027, USA}

$^{67}$ {Department of Physics and Astronomy, University of
  Manitoba, MB R3T 2N2, Canada} 

$^{68}$ {Department of Physics and Mathematics, Aoyama Gakuin
  University, Kanagawa 252-5258, Japan} 

$^{69}$ {Astronomical Observatory, Jagiellonian University,
   30-244 Krak\'ow, Poland}

$^{70}$ {Institute of Space-Earth Environmental Research,
  Nagoya University, Aichi 464-8601, Japan} 

$^{71}$ {Advanced Medical Instrumentation Unit, Okinawa
  Institute of Science and Technology Graduate University,
  Okinawa 904-0495, Japan} 

$^{72}$ {Department of Physics, Saitama University, Saitama
  338-8570, Japan} 

$^{73}$ {Science Education, Faculty of Education, Shizuoka
  University, Shizuoka 422-8529, Japan} 

$^{74}$ {Faculty of Health Science, Nihon Fukushi University, Aichi
  475-0012, Japan} 

$^{75}$ {MTA-E\"otv\"os University Lend\"ulet Hot Universe Research
  Group, Budapest 1117, Hungary}

$^{76}$ {Department of Theoretical Physics and Astrophysics,
  Faculty of Science, Masaryk University, 611 37 Brno, Czech Republic}

$^{77}$ {Department of Physics and Astronomy, Johns Hopkins
  University, Baltimore, MD 21218, USA} 

}

\end{document}